\documentclass[%
 reprint,
 amsmath,amssymb,
 aps,
]{revtex4-2}
\usepackage{graphicx}% Include figure files
\usepackage{dcolumn}% Align table columns on decimal point
\usepackage{bm}% bold math
\usepackage{multirow}
\usepackage{hhline}
\usepackage{amsmath}

\begin{document}

\preprint{APS/123-QED}

\title{Large modulation of thermal transport in 2D semimetal triphosphides by doping$-$induced electron$-$phonon coupling}

\email{shenghong.ju@sjtu.edu.cn}

\author{Yongchao Rao$^a$, C. Y. Zhao$^a$, Lei Shen$^b$, and Shenghong Ju$^{a,*}$}
\affiliation{$^a$China-UK Low Carbon College, Shanghai Jiao Tong University, Shanghai 201306, China\\
$^b$Department of Mechanical Engineering, National University of Singapore, 117575, Singapore}

\date{\today}
\begin{abstract}
Recent studies demonstrate that novel 2D triphosphides semiconductors possess high carrier mobility and promising thermoelectric performance, while the carrier transport behaviors in 2D semimetal triphosphides have never been elucidated before. Herein, using the first$-$principles calculations and Boltzmann transport theory, we reveal that the electron$-$phonon coupling can be significant and thus greatly inhibits the electron and phonon transport in electron$-$doped BP$_{\rm 3}$ and CP$_{\rm 3}$. The intrinsic heat transport capacity of flexural acoustic phonon modes in the wrinkle structure is largely suppressed arising from the strong out$-$of$-$plane phonon scatterings, leading to the low phonon thermal conductivity of 1.36 and 5.33~W/(mK) for BP$_{\rm 3}$ and CP$_{\rm 3}$ at room temperature, and at high doping level, the enhanced scattering from electron diminishes the phonon thermal conductivity by 71 \% and 54 \% for BP$_{\rm 3}$ and CP$_{\rm 3}$, respectively. Instead, electron thermal conductivity shows nonmonotonic variations with the increase of doping concentration, stemming from the competition between electron$-$phonon scattering rates and electron group velocity. It is worth noting that the heavy$-$doping effect induced strong scattering from phonon largely suppresses the electron transport and reduces electron thermal conductivity to the magnitude of phonon thermal conductivity. This work sheds light on the electron and phonon transport properties in semimetal triphosphides monolayer and provides an efficient avenue for the modulation of carrier transport by doping$-$induced electron$-$phonon coupling effect.
\end{abstract}
\maketitle

\section{Introduction}
Thermal transport underlies the operation of modern devices from thermoelectric modules to thermal barrier systems to heat$-$management systems\cite{1,2,3,4}. As one of the most appealing fundamental physical problems, efficiently manipulating thermal transport in materials show enormous practical implications. The interaction between two thermal transport carriers, electron and phonon, plays important role in a variety of physical phenomena, such as the electron pairing$-$induced quantum phenomenon of superconductivity\cite{5}, Raman spectra\cite{6,7}, and photoemission spectra\cite{8}. With the improved operability in electron heavily doping, the impact of doping$-$induced electron$-$phonon coupling (EPC) on thermal transport has been revealed to have a non$-$negligible effect on phonon thermal conductivity ($k$$_{\rm ph}$) above room temperature. For example, Liao $et~al$.\cite{9} found an unexpected and significant reduction of the $k$$_{\rm ph}$ in hole$-$doped silicon can reach up to 45 \% as the carrier concentration goes to 10$^{21}$~cm$^{-3}$ using a first$-$principles study, which was subsequently verified via the modified transient thermal grating technique\cite{10}. This inspires successive studies of EPC effect on phonon transport in 3C$-$SiC\cite{11}, wurtzite ZoO\cite{12}, and metals\cite{13}. With the rapid development of nanotechnology, 2D atomic crystals have seen surging interest in a wide range of applications, especially in electronic transport devices. In such systems, two$-$dimensionality allows precise control of the carrier density by a gate\cite{14,15}. So, understanding the doping$-$tuned EPC effect on the dynamics of charge carriers in 2D materials has also received considerable attention\cite{16,17,18,19,20}. Besides, due to the intrinsic enough density of states near the Fermi energy, the strong EPC is more likely to occur in 2D metallic materials\cite{21}.

As one novel type of 2D material, triphosphides monolayer composed of phosphorus and selective elements from Group III, IV, and V, can be easily exfoliated from layered materials. Experimentally, the typical layered materials GeP$_{\rm 3}$ and SnP$_{\rm 3}$ possessing puckered honeycomb threefold coordinated structures were first synthesized in the 1970s\cite{22,23,24}. Subsequent theoretical studies reported the small cleavage energy of the type of 2D triphosphides, suggesting exfoliation of bulk materials as viable means for the preparation of mono$-$ and few$-$layer materials\cite{25,26,27,28}. Importantly, the electronic characteristics’ dependence on the number of layers and high carrier mobility similar to phosphorene intrigues researchers to further study its potential application in flexible solar cells\cite{25,26}, ion batteries\cite{29,30}, optoelectronic\cite{26}, and thermoelectric devices\cite{31,32,33}. More recent studies have demonstrated that several 2D triphosphides show the high thermoelectric figure of merit due to their low $k$$_{\rm ph}$\cite{31}. Such low $k$$_{\rm ph}$ can be attributed to low acoustic group velocity and strong phonon$-$phonon (ph$-$ph) scattering. In other words, unlike the graphene$-$like plane structures, the contribution of ZA mode to the in$-$plane thermal conductivity in puckered structures is largely suppressed due to the strong ph$-$ph scattering along the out$-$of$-$plane directions. Moreover, the researches on CaP$_{\rm 3}$\cite{34}, GeP$_{\rm 3}$\cite{35}, and AsP$_{\rm 3}$\cite{36} also elucidate that they are hopeful alternatives for application in thermoelectric technologies mainly originating from their low $k$$_{\rm ph}$.

Triggered by the fascinating findings, an obvious question arises whether the elements of the short period in the periodic table can exit in 2D triphosphides, if they can exit, then what properties and applications can be expected? In this regard, researchers have made efforts to broaden the 2D triphosphides family using theoretical methods. Kar $et~al.$\cite{37} systematically studied the thermal, mechanical, and dynamical stabilities of 2D CP$_{\rm 3}$, and predicted a route to synthesize CP$_{\rm 3}$ monolayer through C atoms doping into blue phosphorene. Notably, the electron delocalization resulting in the metallic behavior of CP$_{\rm 3}$ monolayer further motivates us to investigate the EPC effect in 2D semimetal triphosphodes. As mentioned, the existence of electron density$-$of$-$states (el$-$DOS) near the fermi level may strengthen the electron$-$phonon (el$-$ph) interaction. Besides, BP$_{\rm 3}$ monolayer analogous to blue phosphorene was also reported by Shojaei and Kang using the CALYPSO\cite{38}. In fact, from the foundational understanding of the interactions between electrons and phonons, great efforts have been made in semimetal W$_{\rm 2}$N$_{\rm 3}$\cite{39} and Beryllonitrene\cite{40} by means of first$-$principles calculations. In terms of the planar structures, these investigations give detailed studies of the transport properties from el$-$ph scattering. However, as for the puckered structures with ZA branches being greatly suppressed, the thermal transport considering the effects of EPC in 2D semimetal BP$_{\rm 3}$ and CP$_{\rm 3}$ remains unexplored.

In this work, we conduct comprehensive first$-$principles calculations to investigate the electron$-$doping induced el$-$ph scattering effect on the thermal transport of 2D semimetal BP$_{\rm 3}$ and CP$_{\rm 3}$. In puckered structures, the strong coupling between acoustic flexural phonons results in low $k$$_{\rm ph}$, and optical branches dominate the phonon transport. At a high doping level, our calculation explicitly shows the magnitude of el$-$ph scattering is comparable with the ph-ph scattering in the high$-$frequency region, thus resulting in the remarkable suppression of phonon transport. The electron thermal conductivity ($k$$_{\rm e}$) displays anomalous variations with the increase of doping concentration due to the competition between el$-$ph scattering rates and electron group velocity. In heavily doped systems, strong scattering of high$-$frequency optical phonons onto electrons gives rise to the large reduction of $k$$_{\rm e}$ to the magnitude of $k$$_{\rm ph}$. The Boltzmann transport equation (BTE) calculations reveal effective ways to tune the thermal transport in semimetal BP$_{\rm 3}$ and CP$_{\rm 3}$ by doping$-$induced el$-$ph scattering.

\section{Computational details}
Density functional theory (DFT) and density functional perturbation theory (DFPT) calculations were carried out with the Quantum Espresso package\cite{41}. The Perdew$-$Burke$-$Ernzerhof exchange and correlation functionals\cite{42} were employed with the projector$-$augmented wave method. A vacuum thickness of 12~{\AA} is introduced in the simulation domain to eliminate the interactions between different layers. A plane wave cutoff was set to be 100~Ry, and a $15\times15\times1$ $k$$-$mesh and a $5\times5\times1$ $q$$-$mesh were employed for electronic and DFPT calculations, respectively. The convergence threshold of electron energy is set to be 10$^{-12}$~Ry. The left panel of Fig. 1 shows the crystal structure of 2D triphosphides, and this phase is similar to blue phosphorene in that two P atoms in the $2\times2$ hexagonal unit cell of the latter are replaced by two B or C atoms. The optimized lattice parameters are 6.53 and 6.24~{\AA} for BP$_{\rm 3}$ and CP$_{\rm 3}$, respectively, which are consistent with previous first$-$principles results (6.50~{\AA} for BP$_{\rm 3}$\cite{38} and 6.22~{\AA} for CP$_{\rm 3}$\cite{37}). By iteratively solving the BTE, the $k$$_{\rm ph}$ tensor can be written as\cite{43}
\begin{equation}
k_{ph}^{\alpha\beta} = \frac{1}{N_qV}\sum_{\lambda}{\hbar}{\omega_\lambda}{\frac{\partial{n_{\lambda}^{0}}}{\partial{T}}}{v_{\lambda}^{\alpha}}{v_{\lambda}^{\beta}}\tau_{\lambda}^{ph}
\end{equation}
Where $N_q$, $V$, $\hbar$ , $\omega_{\lambda}$, and $n_{\lambda}$ are the number of $q$$-$points in the first Brillouin zone, the volume of the primitive cell, reduced Planck constant, frequency of phonon branch mode $\lambda$ = $(q,v)$ with polarization $v$ and wave vector $q$, and Bose$-$Einstein distribution function, respectively. $v_{\lambda}^{\alpha}$ is the projection of the phonon group velocity along the direction $\alpha$. The relaxation time, $\tau_{\lambda}$, considering both ph$-$ph scattering and phonon$-$electron (ph$-$el) scattering can be determined by the Matthiessen's rule\cite{43}, $1/\tau_{\lambda}$ = $1/\tau_{\lambda, pp}$ + $1/\tau_{\lambda, pe}$, where $1/\tau_{\lambda, pp}$ and $1/\tau_{\lambda, pe}$ are the three$-$phonon (3ph) scattering rates and ph$-$el scattering rate, respectively.

For the 3ph scattering process, we adopted the finite difference method as implemented by thirdorder.py\cite{44} to calculate third$-$order interatomic force constants (3rd IFCs) using a $3\times3\times1$ supercell. The $18^{\rm {th}}$ nearest neighbors were included for the calculations of 3rd IFCs, and then, the 3ph scattering rates and intrinsic $k$$_{\rm ph}$ were calculated using ShengBTE\cite{44} with $q$$-$points of $60\times60\times1$ sampling the Brillouin zone. The convergence of $k$$_{\rm ph}$ with respect to IFCs cutoffs and $q$$-$points sampled in the Brillouin zone was carefully discussed (see Fig. S1). $1/\tau_{\lambda, pe}$ was given by the imaginary part of the phonon self$-$energy under a complete EPC calculation based on the DFPT and Wannier interpolation techniques. Coarse grids of $15\times15\times1$ $k$$-$mesh and $5\times5\times1$ $q$$-$mesh were interpolated into the dense meshes of $180\times180\times1$ $k$$-$mesh and $60\times60\times1$ $q$$-$mesh through the maximally localized Wannier functions as implemented in the EPW code\cite{45,46}. The Dirac $\delta$ functions for electrons and phonons were smeared out by a Gaussian function with widths of 20 and 0.2~meV, respectively. After obtaining the total phonon scattering $1/\tau_{\lambda}$, $k$$_{\rm ph}$ considering the ph$-$el effect was computed by solving equation (1) with an iterative schema. To deal with the thermal transport value of the 2D system, the effective thickness along the vacuum axis is determined by the sum of the thickness of the monolayer sheet and the van der Waals radius of the top and bottom atoms.

Based on the electronic band structure obtained from Wannier interpolation techniques, the $k$$_{\rm e}$, and electrical conductivity ($\sigma$) are evaluated by using the electronic BTE and Onsager relations\cite{47}
\begin{equation}
\begin{split}
k_{e}(\mu,T) =& \frac{1}{N_kV}\sum_{n, k}{-\frac{(\epsilon_{nk}-\mu)}{T}}{v_{nk}^2}\tau_{nk}(\mu, T)\\
&\times\frac{\partial{\emph{f$_{ED}$}(\epsilon_{nk},\mu,T)}}{\partial{\epsilon}}-{TS^2(\mu,T)\sigma(\mu,T)}\\
\end{split}
\end{equation}

\begin{equation}
\sigma(\mu,T) = \frac{1}{N_kV}\sum_{n, k}{-e^2}{v_{nk}^2}\tau_{nk}(\mu, T)\frac{\partial{\emph{f$_{ED}$}(\epsilon_{nk},\mu,T)}}{\partial{\epsilon}}
\end{equation}
Where $\mu$, $T$, $N_k$, and $e$ are chemical potential, temperature, the total number of $k$$-$points, and elementary charge, respectively. $\epsilon_{nk}$ and $v_{nk}$ are the energy eigenvalue and group velocity of band index $n$ at state $k$, $f_{ED}$ is the Fermi$-$Dirac distribution function, and $S$ is the Seebeck coefficient. The electron relaxation time $\tau_{nk}$, limited by el$-$ph scattering, can be obtained from the imaginary part of electron self$-$energy.

\section{Results and Discussion}
In the right panel of Fig. 1, we characterize bonding with the electron localization function (ELF). Clearly, ELF is the largest at the center of all P$-$P and P$-$B/C bonds, indicating the covalent bonding characteristic in the BP$_{\rm 3}$ and CP$_{\rm 3}$. Such strong in$-$plane bonds are in analogy with the covalent bond$-$driven metallic responsible for the high critical temperature ($T_c$) in the MgB$_{\rm 2}$ superconductor\cite{48}. Moreover, these bonds are solely responsible for the bands crossing the Fermi level, becoming ideal candidates with strong el$-$ph interactions.

\begin{figure}[h]
  \centering
  \includegraphics[width=8.6cm]{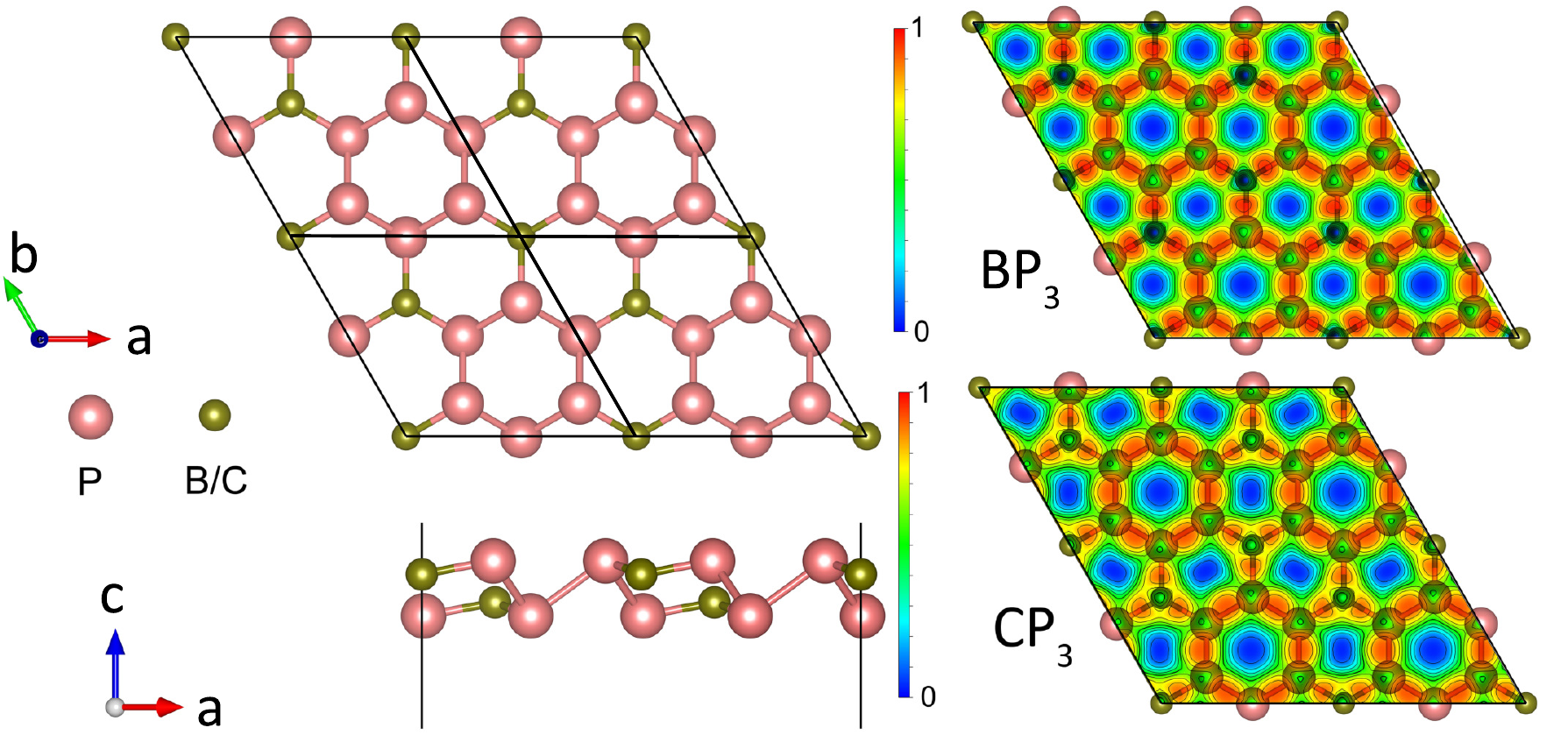}
  \caption{Top and side views of BP$_{\rm 3}$ and CP$_{\rm 3}$ are shown in the left panel. The primitive cell consists of 2 B/C atoms and 6 P atoms. The ELF parallel to the (001) plane is shown in the right panel.}
\end{figure}

\begin{figure*}
  \centering
  \includegraphics[width=15cm]{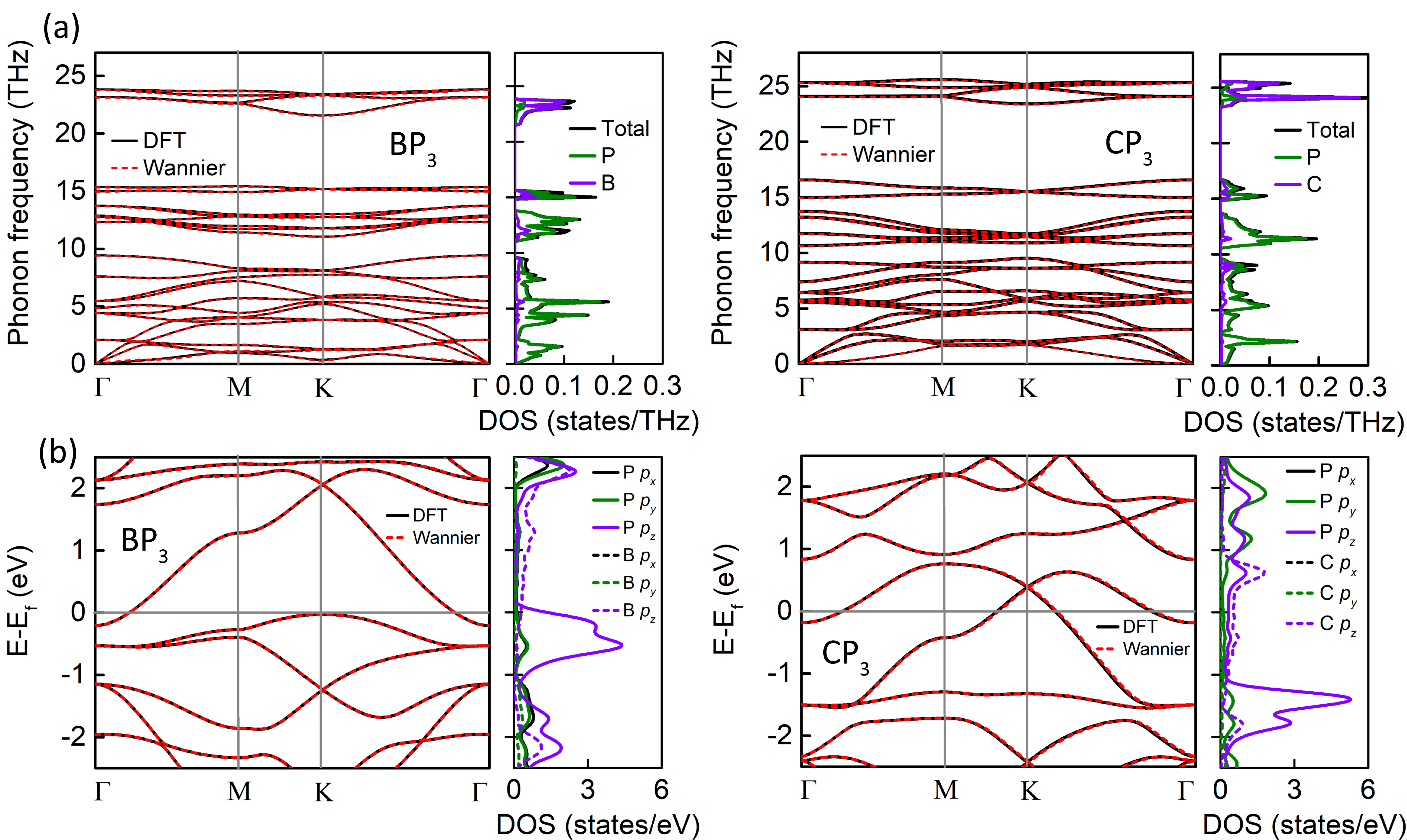}
  \caption{(a) Phonon dispersion, and (b) Electronic band structure with the corresponding DOS for BP$_{\rm 3}$ and CP$_{\rm 3}$, respectively. The black and red lines represent the results calculated by DFT and Wannier interpolation technique. The electron energy is set to the Fermi energy. }
\end{figure*}

Figure 2(a) depicts the phonon dispersions and phonon density$-$of$-$states (DOS) of BP$_{\rm 3}$ and CP$_{\rm 3}$, and both acoustic and optical modes are primarily contributed by P elements except for the extremely high$-$frequency optical modes. Two triphosphides show dynamic stability since there are no imaginary phonon modes. And the thermal stabilities of 2D BP$_{\rm 3}$ and CP$_{\rm 3}$ are confirmed by performing AIMD simulations shown in Figs. S4 and S5. Attentionally, the highest frequency modes of BP$_{\rm 3}$ and CP$_{\rm 3}$ appear around 25~THz, which is higher than those of metal triphosphides (14~THz)\cite{31}, Silicene (17~THz)\cite{49}, MoS$_{\rm 2}$ (14~THz)\cite{50}, phosphorene (13.5~THz)\cite{51}, indicating the mechanical robustness of the two triphosphides. Besides, the low$-$frequency acoustic branches are much lower (\textless~2.5~THz) than the common 2D materials, including graphene\cite{52}, BN\cite{53}, transition metal disulfides\cite{50,54}, and phosphorene\cite{51}, implying that the phonon harmonic vibrations of 2D triphosphides are extremely weak, as well as the localization of phonon branches with considerably high 3ph scattering strength, which is a signature of low $k$$_{\rm ph}$. Fig. 2(b) plots the electronic band structures and projected el$-$DOS. Both BP$_{\rm 3}$ and CP$_{\rm 3}$ are semimetals, in which electron and hole pockets coexist on the Fermi surface, unlike the semiconductor characteristic in 2D metal triphosphides. For more insights, we can see the bands crossing the Fermi energy level mainly consist of the $p_{z}$ orbitals of P atoms and the $p_{z}$ orbital of B/C atoms. It is concluded that semimetal behaviors are caused by $\pi$ types of interaction induced electron delocalization.

\begin{figure*}
  \centering
  \includegraphics[width=14cm]{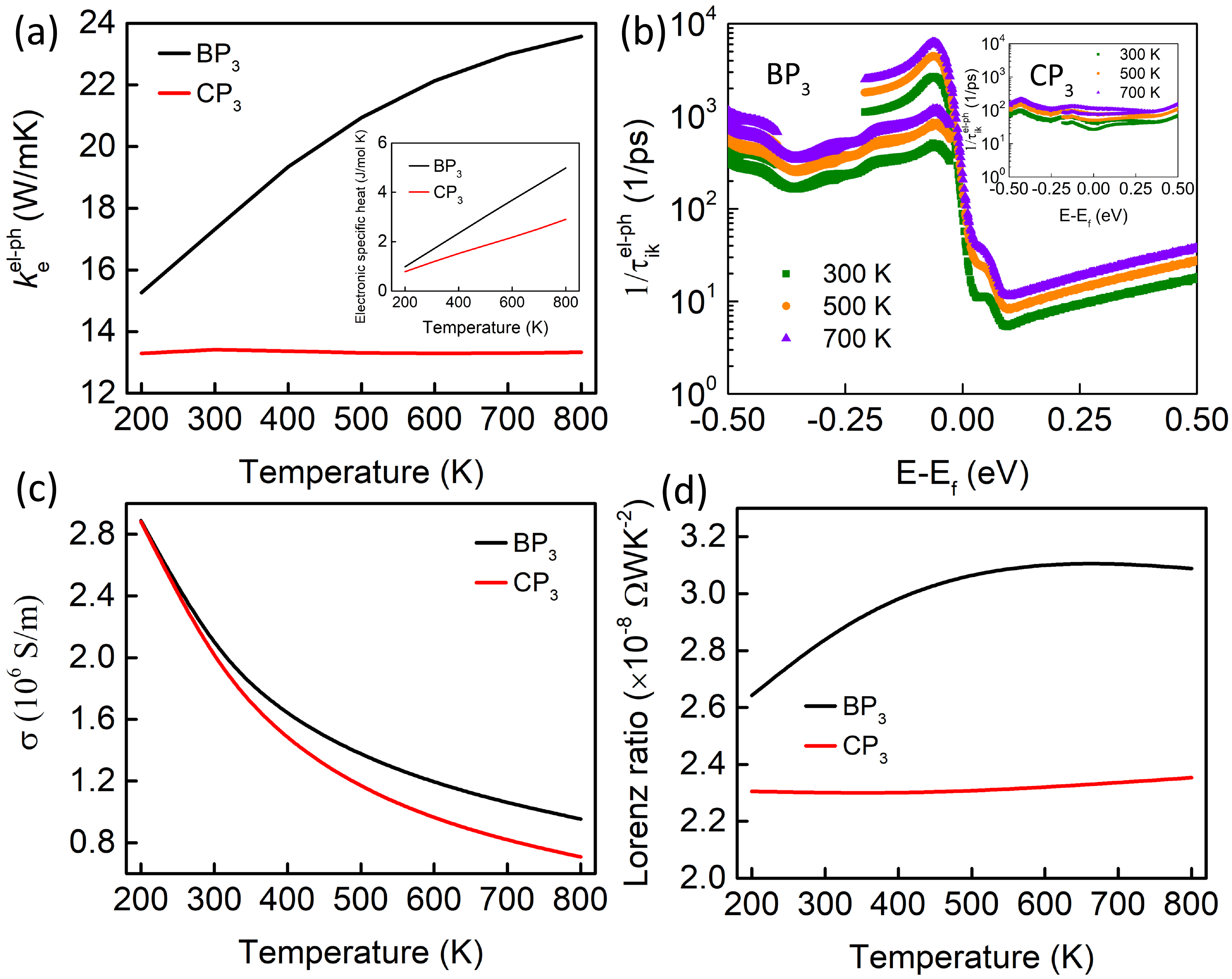}
  \caption{ (a) Electron thermal conductivity $k$$_{\rm e}$, (b) Electron scattering rates 1/{$\tau_{ik}$} at 300, 500, and 700~K, (c) Electrical conductivity $\sigma$, and (d) Lorenz ratio $versus$ $T$ for BP$_{\rm 3}$ and CP$_{\rm 3}$, respectively. The insert in (a) represents the electronic specific heat vary with $T$. }
\end{figure*}

\begin{figure*}
  \centering
  \includegraphics[width=14cm]{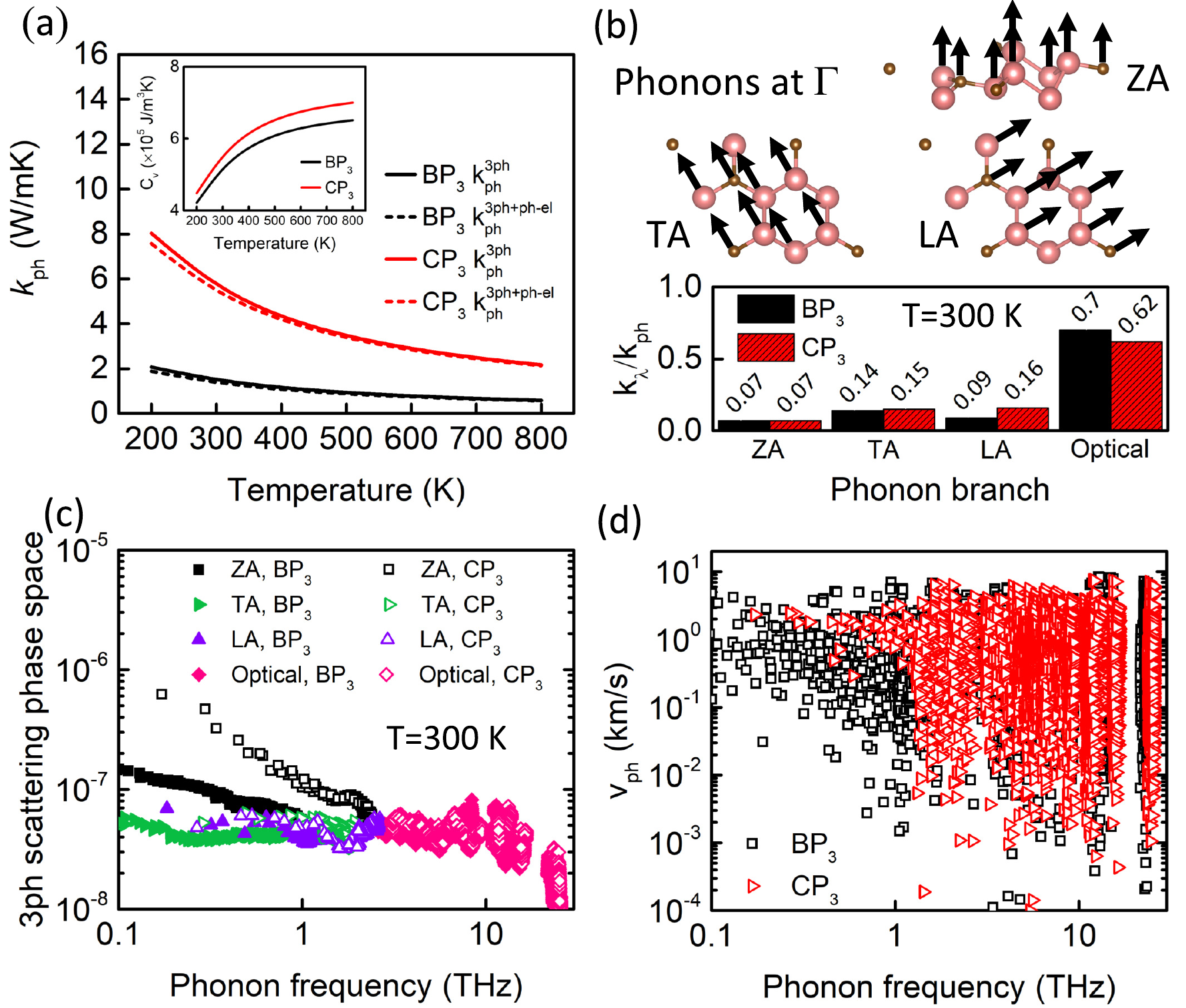}
  \caption{(a) The $k$$_{\rm ph}^{\rm 3ph}$ and $k$$_{\rm ph}^{\rm 3ph+ph-el}$ $versus$ $T$, the inset is the variation of volumetric heat capacity  with $T$. (b) Variational pattern of ZA, TA, and LA phonon modes near the $\Gamma$ point, the bar chart is the ratio of $k$$_{\rm \lambda}$/$k$$_{\rm ph}$ at $T$ = 300~K. (c) Phase space of 3ph scattering processes at 300~K. (d) Phonon group velocity for BP$_{\rm 3}$ and CP$_{\rm 3}$.}
\end{figure*}

To understand the thermal transport mechanisms in the semimetal BP$_{\rm 3}$ and CP$_{\rm 3}$ monolayer, we first examine the separated electron and phonon contributions to thermal conductivity without considering carrier doping. As shown in Fig. 3(a), the $k$$_{\rm e}$ at 300~K are 17.40 and 13.41~W/(mK) for BP$_{\rm 3}$ and CP$_{\rm 3}$, respectively. In semimetal triphosphides, the small density of states at the Fermi level plays non$-$negligible roles in the electronic contribution to heat conduction. Obviously, the $k$$_{\rm e}$ of CP$_{\rm 3}$ is almost temperature$-$independent, ascribing to the complementary contributions between the decrease of electron relaxation time and the increase of electronic heat capacity with $T$. However, the value of BP$_{\rm 3}$ increases with $T$. As plotted in Fig. 3(b), the electron scattering rates for BP$_{\rm 3}$ near the Fermi level show a slight increase with the $T$ increasing, which is incomparable with the increasing amplitude in CP$_{\rm 3}$. Besides, the increase rate in electronic heat capacity for BP$_{\rm 3}$ is higher than that of CP$_{\rm 3}$ (inset in Fig. 1 (a)). These features result in the difference in $k$$_{\rm e}$ as a function of $T$ between BP$_{\rm 3}$ and CP$_{\rm 3}$. We further calculate the $\sigma$, as illustrated in Fig. 3 (c). The magnitude of $\sigma$ is on the order of 10$^{\rm 6}$~S/m, which is significantly smaller than that of typical metals\cite{16,55} and 2D carbon allotropes\cite{56} also due to the lower el$-$DOS near the Fermi level.

The theoretical methods can accurately obtain $k$$_{\rm e}$ and $k$$_{\rm ph}$, while the experimental techniques are difficult to measure separately. Although some methods have been developed for direct $k$$_{\rm ph}$ measurements, including the magnetothermal technique\cite{57} and alloying technique\cite{58}, they are limited to the complicated operation process and extremely low temperatures. Benefiting from the easy measurement of electrical conductivity, the approximation, $k_{\rm ph} = k_{\rm total} - L{\sigma}T$, is employed to obtain $k$$_{\rm ph}$. Here, the Lorenz ratio is assumed to be the Sommerfeld value $L_0$ = 2.44 × 10$^{-8}$~$\Omega$WK$^{-2}$\cite{59}. This approximation is inaccurate when the temperature is far lower than the Debye temperature since the $L$ is a chemical$-$potential and temperature$-$dependent quantity, due to the fact that the el$-$ph scattering rate in the vicinity of Fermi level is not a constant with respect to the electron energy and temperature. Hence, we also plot the $L$ as a function of $T$ in Fig 3 (d). Both calculated $L$ in BP$_{\rm 3}$ and CP$_{\rm 3}$ show a similar trend as $k$$_{\rm e}$ $versus$ $T$. The value in BP$_{\rm 3}$ is bigger than the Sommerfeld value while the value in CP$_{\rm 3}$ is smaller. Both $L$ increase with $T$, which is in agreement with the Bloch$-$Grüneisen theory\cite{60}.

\begin{figure*}
  \centering
  \includegraphics[width=14cm]{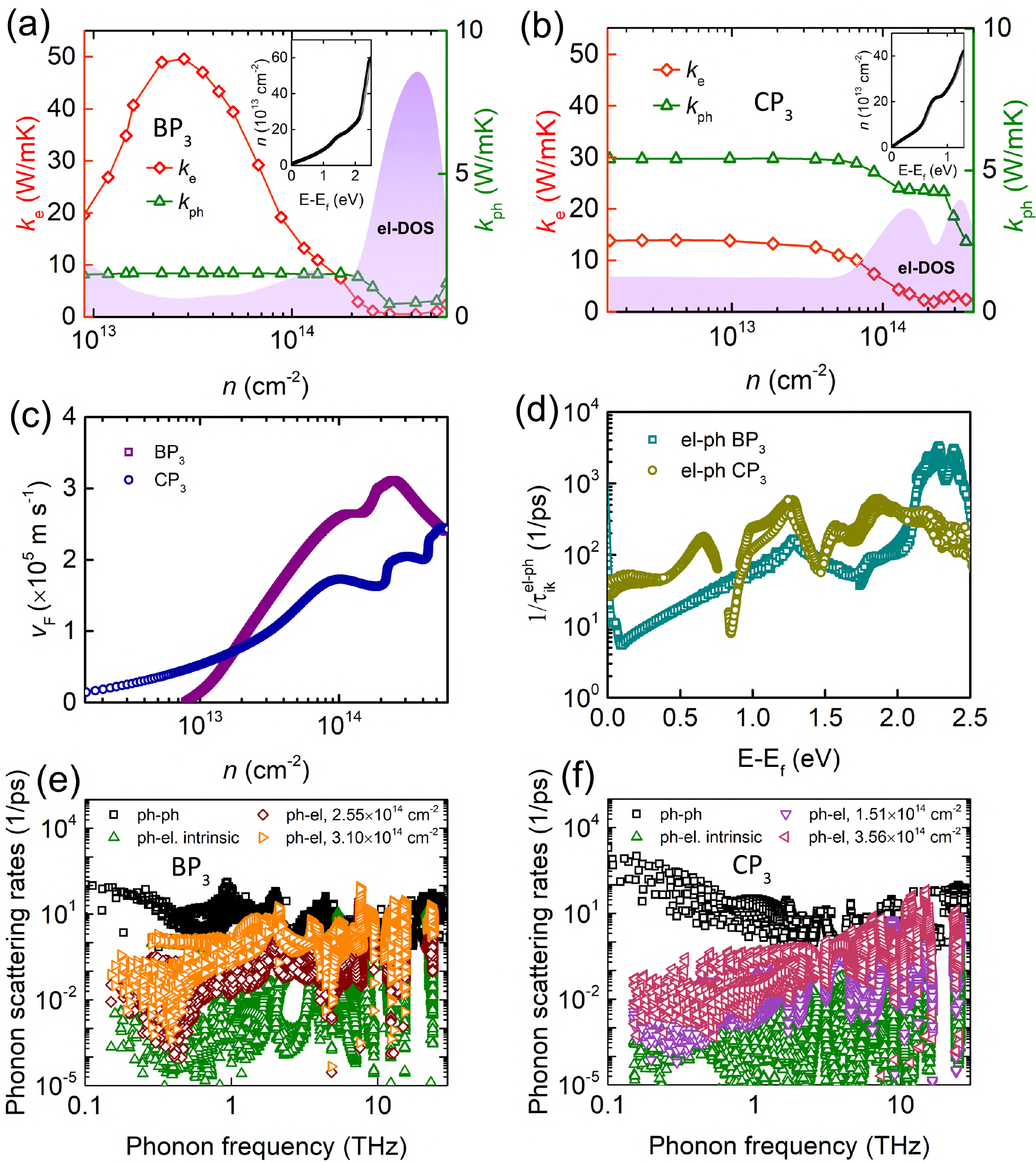}
  \caption{The $k$$_{\rm e}$ and $k$$_{\rm ph}$ as a function of $n$ for electron$-$doped (a) BP$_{\rm 3}$ and (b) CP$_{\rm 3}$. The purple shaded area is the el$-$DOS, and the corresponding insert is the variation of $n$ with shifted Fermi energy. (c) $v_F$ and (d) el$-$ph scattering rates vary with $n$ and shifted Fermi energy, respectively. The calculated scattering rates of 3ph and ph$-$el for intrinsic and electron$-$doped with different $n$ in (e) BP$_{\rm 3}$ and (f) CP$_{\rm 3}$ at T = 300~K.   }
\end{figure*}

Next, we discuss the phonon contribution to thermal conductivity. Figure 4 (a) explicitly shows the temperature$-$dependent $k$$_{\rm ph}$ of the intrinsic systems. The $k$$_{\rm ph}$ follows the normal $T^{\rm -1}$ trend generally. As expected, the $k$$_{\rm ph}$ are only 1.48 and 5.60~W/(mK) at 300~K without considering ph$-$el scattering for BP$_{\rm 3}$ and CP$_{\rm 3}$, respectively. For phosphorene$-$like compounds with puckered structures, the contribution of ZA mode for $k$$_{\rm ph}$ is relatively low due to the strong ph$-$ph scattering perpendicular to the plane direction. The contribution from the ZA mode shows differences compared with the atomically flat plane, such as representative graphene\cite{13} and BN\cite{61}, in which the ZA mode contributes to the most of $k$$_{\rm ph}$. For quantified insights, Fig. 4 (b) depicts the contributions of different phonon branches to $k$$_{\rm ph}$ at 300~K. The acoustic modes contributing to the $k$$_{\rm ph}$ are indeed smaller than optical modes, and the heat conductance contributed from ZA mode is the least. The above discussed results can be understood by plotting the intrinsic ph$-$ph scattering rates in Figs. 5 (e) and (f). The phonon modes in the low$-$frequency range present the larger ph$-$ph scattering rates, indicating the stronger ph$-$ph scattering strength and lower phonon lifetime, thus predominately limiting the finite thermal conductivity of acoustic modes. Moreover, the strength of 3ph scattering is also reflected in the inverse of phonon phase space. We plot the scattering phase space of ZA, TA, LA, and Optical branches in Fig. 4(c), confirming the abundant scattering channels of the acoustic modes, especially for the ZA modes. Additionally, the insert in Figs. 4 (a) and (d) depict the temperature$-$dependent volumetric heat capacity and phonon group velocity, respectively. The larger volumetric heat capacity in CP$_{\rm 3}$ is one of the key reasons why its $k$$_{\rm ph}$ is higher than BP$_{\rm 3}$. Owing to the much$-$alike phonon dispersions between BP$_{\rm 3}$ and CP$_{\rm 3}$, the group velocities of these two systems show inconspicuous differences. Furthermore, the $k$$_{\rm ph}$ does not show significant reduction after considering the ph$-$el scattering, which means the extremely weak ph$-$el scattering in 2D triphosphides, unlike the metallic carbides\cite{62} and nitrides\cite{63} systems. To validate the inference, the ph$-$ph and ph$-$el scattering rates are comparatively plotted in Figs. 5 (e) and (f) for BP$_{\rm 3}$ and CP$_{\rm 3}$, respectively. We can see that the ph$-$ph scattering rates are obviously larger than that of ph$-$el scattering rates in the whole frequency ranges, which means the phonon lifetime is dominated by ph$-$ph scattering in the undoped triphosphides.

The atomic$-$thick structure endows 2D materials with a greater possibility to manipulate their properties by doping. We next investigate how electron doping can be used to tune the el$-$ph interactions and then rationally modulate the thermal transport performance. The $k$$_{\rm e}$ and $k$$_{\rm ph}$ for both compounds at 300~K varying with electron doping concentration ($n$) are shown in Figs. 5(a) and (b). The $k$$_{\rm e}$ of BP$_{\rm 3}$ exhibits an unusual nonmonotonic dependence on $n$, while the $k$$_{\rm e}$ of CP$_{\rm 3}$ is insensitive with
$n$ ($\textless$ 10$^{13}$~cm$^{\rm -2}$), and begins to remarkedly decrease beyond 10$^{13}$~cm$^{\rm -2}$. Compared with the intrinsic system, the $k_{\rm e}$ of BP$_{\rm 3}$ increases by up to 185\% (from 17.40 to 49.58~W/(mK)) at $n$ = 2.85$\times$10$^{13}$~cm$^{\rm -2}$, while further increasing $n$ up to 4.20$\times$10$^{14}$~cm$^{\rm -2}$, the $k$$_{\rm e}$ reduces to 0.51~W/(mK). The $k$$_{\rm e}$ of CP$_{\rm 3}$ reduces from 13.41 to 1.94~W/(mK) at $n$ = 2.18$\times$10$^{14}$~cm$^{\rm -2}$. The concentration$-$dependent variation trend of the $k$$_{\rm e}$ can be approximately captured by Drude’s free electron model, $k_e = {\pi}^2k_B^2TN_Fv_F{\tau}_e/3$.\cite{59} Here, $N_F$ is the Fermi DOS highlighted as purple shades in Figs. 5(a) and (b). $v_F$ is the Fermi velocity and ${\tau}_e$ is the electron relaxation time, both of which are plotted in Figs. 5(c) and (d), respectively. At a low doping level, the increased tendency of $k$$_{\rm e}$ in BP$_{\rm 3}$ mainly results from the sharply lowered el$-$ph scattering rates, as the lowered el$-$DOS is fully compensated by the enhanced $v_F$ and $\tau_e$. When continuing to increase $n$ ($\textgreater$ 2.85$\times$10$^{13}$~cm$^{\rm -2}$), gradually enhanced el$-$ph interactions result in a minimum $k$$_{\rm e}$ in spite of surging el$-$DOS and increased $v_F$. The $k$$_{\rm e}$ of CP$_{\rm 3}$ showing an insensitive trend with $n$ at a low doping level ($\textless$ 10$^{13}$~cm$^{\rm -2}$) stems from the compensated relationship by the enhanced $v_F$ and decreased $\tau_e$. Further increasing $n$ gives rise to remarkably enhanced interaction between electron and phonon, and eventually suppressed the electron thermal transport capacity. For both systems, the EPC strength generally increases with $n$ increasing. In addition, the electron dominates the overall thermal transport over a wide range ($n$ $\textless$ 1.05$\times$10$^{14}$~cm$^{\rm -2}$), however, when beyond the doping level, the sharp suppression in $k$$_{\rm e}$ by largely enhanced EPC leads to $k$$_{\rm ph}$ surpassing $k$$_{\rm e}$.

The $k$$_{\rm ph}$ in both BP$_{\rm 3}$ and CP$_{\rm 3}$ shows a monotonous decrease, which becomes more remarkable when the doping level reaches to the vicinity of high el$-$DOS. The reduction of $k$$_{\rm ph}$ in BP$_{\rm 3}$ and CP$_{\rm 3}$ can reach 71\% at $n$ = 3.10$\times$10$^{14}$~cm$^{\rm -2}$ and 54\% at n = 3.56$\times$10$^{14}$~cm$^{\rm -2}$, respectively. Similar behavior can be observed in the heavily doped silicon\cite{9} and 2D Dirac silicene\cite{12} and Beryllonitrene\cite{40}. We plot the ph$-$el scattering rates in Fig.5 (e) and (f) to quantitatively compare the importance of ph$-$ph and ph$-$el scattering evens. It can be clearly seen that the ph$-$ph scattering rates are much higher than ph$-$el scattering rates in lightly doped systems, which then confirms $k$$_{\rm ph}$ insensitive dependence with $n$ at the low doping level. However, the ph$-$el scattering rate becomes comparable at the high doping level. In particular, the high$-$frequency phonons show remarkable interaction with electrons, which will largely suppress the heat transport capacity of optical modes. As mentioned earlier, optical modes dominate the phonon thermal transport in puckered BP$_{\rm 3}$ and CP$_{\rm 3}$. So, the decreased trend of $k$$_{\rm ph}$ at high $n$ is the result of the enhanced electrons scattering onto phonons. In total, the EPC can significantly reduce the overall thermal conductivity. This anomalous thermal transport regime by doping$-$induced strong EPC provides a theoretical basis for experimental detection of the EPC effect on thermal conductivity.

\section{Conclusion}
By employing the first$-$principles calculations, we have systematically predicted the $k$$_{\rm e}$ and $k$$_{\rm ph}$ of 2D semimetal BP$_{\rm 3}$ and CP$_{\rm 3}$. It has been demonstrated the large modulation of thermal conductivity via the electron doping$-$induced EPC effect. The suppression of thermal transport of ZA modes in puckered structures results in low $k$$_{\rm ph}$, and electrons dominate the thermal transport. In addition, it is found that the high$-$frequency optical phonon modes are susceptible to scatter onto electrons. When increasing the doping concentration, the $k$$_{\rm e}$ in BP$_{\rm 3}$ and CP$_{\rm 3}$ show different variation trends due to the competition between the $v_F$ and $\tau_e$. However, the $k$$_{\rm ph}$ in both systems undergoes a monotonic decrease because of the gradually enhanced ph$-$el interaction with increased $n$ and their reductions become significant only at a high doping level ($n$ $\textgreater$ 10$^{14}$~cm$^{\rm -2}$). Furthermore, upon $n$ $\textgreater$ 1.05$\times$10$^{14}$~cm$^{\rm -2}$, the $k$$_{\rm e}$ of both compounds begins to lower than $k$$_{\rm ph}$ due to sharply enhanced el$-$ph interaction. Our works shed light on the intriguing thermal conductivity modulation mechanism in 2D semimetal triphosphides by doping$-$induced EPC, and the large modulation range of thermal conductivity indicates its potential application in thermal switching devices.

\section*{Supplementary materials}
See supplementary materials for the convergence of $q$$-$points and IFCs cutoff in phonon thermal conductivity calculations, validation of the Wannier interpolation, and thermal stability test by $ab~initio$ molecular dynamics simulations.

\section*{Acknowledgments}
This work was supported by the National Natural Science Foundation of China (No. 52006134) and the Shanghai Key Fundamental Research via Grant No. 21JC1403300. The computations in this paper were run on the $\pi$ 2.0 cluster supported by the Center for High Performance Computing at Shanghai Jiao Tong University.

\end{document}